\documentstyle[12pt,a4wide]{article}
\include{epsf}

\begin{document}

\Large

\noindent
{\bf Frequency locking to a high-finesse Fabry-Perot cavity of a 
frequency doubled Nd:YAG laser used as the optical phase modulator}

\vspace{1cm}
\normalsize
M. Bregant, G. Cantatore, F. Della Valle

\small
{\it  Dipartimento di Fisica and INFN, via Valerio 2, 
34127 Trieste, Italy}

\normalsize
G. Ruoso

\small
{\it  INFN, Laboratori Nazionali di Legnaro, via Romea 4, 
35020 Legnaro, Italy}

G. Zavattini

\small
{\it Dipartimento di Fisica and INFN, via del Paradiso 
12,  Ferrara, Italy}

\normalsize

\abstract{We report on the frequency locking of a frequency doubled 
Nd:YAG 
laser to a 45\,000 finesse, 87-cm-long, Fabry-Perot cavity using a 
modified form of the 
Pound-Drever-Hall technique.  Necessary signals, such as light phase 
modulation and frequency correction feedback, are fed direcly to the 
infrared  pump laser. This is sufficient to achieve a stable locking 
of the 532 nm visible beam to the cavity, also showing that the 
doubling process does not degrade laser performances.}

\newpage

\section{Introduction}

The use of high finesse Fabry-Perot (FP) cavities has become 
widespread in the last decade, 
especially due to the availability of 
very  high 
reflectivity mirrors. FP resonators play a crucial role in several 
physical 
fields, such as QED and vacuum structure measurements with optical 
techniques$^{1}$, gravitational wave detection with 
interferometers$^{2}$ and resonant bars$^{3}$, and 
metrology. Their use for $H^{0}$ stripping in high intensity proton 
drivers with $H^{-}$ injection has recently been suggested$^{4}$.
 The resonant operating condition of these devices is 
achieved by 
frequency locking a laser to one of the fundamental modes of the 
cavity. 
The Pound-Drever-Hall technique$^{5}$ is the most common method 
used to 
accomplish such a locking: the cavity instantaneous frequency is 
sensed by putting frequency-modulation (FM) sidebands on the cavity 
input beam, and a correction signal is then extracted and applied  to 
change accordingly the laser frequency. Tuneable lasers have
proved to be the best suited to implement such a scheme, and the Non 
Planar Ring Oscillator  (NPRO) solid state diode-pumped Nd:YAG laser 
is probably the most common type of laser used in such 
applications. This device 
allows for two different types of modulation:  a "slow" one 
(bandwidth 
$\sim$ 1 Hz, dynamic range of several GHz), and a "fast" one 
(bandwidth $\sim$ 100 kHz, dynamic range $\sim$ 200 MHz). Our group 
has developed an original locking scheme, in which  FM sidebands at  
frequencies $>$ 500 kHz are generated using the laser itself, and the 
method was succesfully applied to a Nd:YAG laser emitting at 1064 
nm$^{6}$. In several applications, such as PVLAS type$^{1}$ 
measurements and $H^{-}$ double stripping$^{4}$, there is an added bonus 
when operating at shorter wavelengths. 
The frequency doubled version of the above mentioned 
laser has become available only recently. In this device the infrared 
(IR) beam exiting the Nd:YAG crystal is fed onto a doubling crystal, 
where, under specific 
conditions, a second beam is generated at twice the original 
frequency. The output beam is a green radiation ($\lambda$ = 532 nm), 
having spectral characteristics which closely match those of the IR 
pump beam. In 
this paper we will show that it is possible to achieve frequency 
locking of the doubled beam with the same good results already 
obtained with 
the IR beam output by a Nd:YAG laser. In particular, we show that it 
is possible to generate 
the FM sidebands on the green output without using an 
external phase modulator. 

\section{Method}

The frequency doubled NPRO laser$^{7}$ is a 
tuneable laser emitting at 532 nm. The green output is generated by a 
single passage of the IR beam, coming from an NPRO Nd:YAG source, 
inside a periodically poled KTP (PP KTP) crystal kept at a stable 
temperature, so as to guarantee the correct phase matching for
second harmonic generation. Tuneability is achieved by changing the 
frequency of the pump laser by means of a temperature control of the 
NPRO crystal (Bandwidth $\sim$ 1 Hz, Dynamic range $\sim$ 60 GHz), 
or of a piezoelectric transducer (Bandwidth $\sim$ 100 kHz, 
Dynamic range $\sim$ 200 MHz) mechanically acting on the laser crystal. 
In order to have the widest possible dynamic range it is compulsory 
to 
use single passage second harmonic generation. This enhances 
stability in the intensity of the green output, but results in a 
lower output power. 

As shown by Cantatore {\it et al.}$^{6}$, phase modulation is equivalent 
to frequency modulation. This allows for the possibility of using the 
laser itself as the optical phase modulator. The phase $\varphi$ of a 
phase-modulated beam can be written as:

\begin{equation}
\varphi (t)=\omega_{0}t + \beta \sin (\Omega_{m}t + \psi)
\end{equation}

where $\omega_{0}$ is the laser angular frequency, $\Omega_{m}$ is 
the angular frequency of modulation and $\beta$ is the modulation 
index. In our setup we are using the piezoelectric actuator on 
the laser head for simultaneous frequency correction and phase 
modulation.

\section{Apparatus}

A schematic drawing of the apparatus is shown in figure 1. 
The laser emits a 100 mW, 532 nm, visible light beam, which is mode 
matched to the 
TEM$_{00}$ mode of a Fabry-Perot cavity by means of a lens. The 
cavity is 87 cm long and is kept under vacuum (total pressure $\leq 
10^{-5}$ mbar); beam input to the vacuum enclosure is possible 
through glass windows. The mirrors forming the FP cavity are high 
reflectivity dielectric mirrors$^{8}$ with an expected 
reflectivity of 0.99997. The 
feedback amplifier is a four-stage integrator circuit with a unity 
gain point that can be chosen in the 20 - 50  kHz interval:
  its transfer function has three poles in the origin and 
one at 0.16 Hz, and the 24 dB/octave slope begins below 7 kHz. A more 
detailed description of the feedback system can be found in Ref. 6. 
The correction signal coming from the feedback amplifier is summed to 
the phase-modulation signal and fed into the laser frequency tuning 
input. 
The possible bandwidth of the feedback loop is limited to the region 
of 
linear behaviour of the laser piezoelectric (PZT) actuator ($\sim$ 
100 kHz), where
it has an actuation coefficient of $\sim$ 2 MHz/V (for the green light 
output). For larger modulation frequencies the response is no longer 
linear, however the 
PZT can still be used at a fixed frequency for phase modulation. 
In order to reduce the unwanted Residual Amplitude Modulation (RAM)
introduced by the phase modulation, one has first to study the 
modulation characteristics of the laser. In fact, as can be seen from 
figure 2, the RAM in the output beam changes with the frequency of 
the 
phase modulation signal. 
A RAM in the beam intensity produces a constant frequency shift 
between the laser and the Fabry-Perot cavity. The working frequency 
is then best set to the value where the RAM reaches a minimum. 
Accordingly, the modulation depth for several values of the 
modulation 
frequency has been measured, and the results are summarized in table 
1. As can be seen from 
figure 3 the 
index of modulation $\beta$ (see Eq. 1) is linear with  the voltage 
$V_{\rm p}$ applied to the laser PZT. In addition, we have also found 
that the ratio 
RAM/$\beta$ 
results to be constant as a function of  $V_{\rm p}$. In the region 
around 660 kHz the RAM has a minimum value, with a ratio RAM/$\beta$ 
of the order $\sim 1 \times 10^{-4}$. This value is about three 
times larger than the value found$^{6}$ using the infrared output of 
an NPRO laser (Lightwave model 124). The difference can probably be 
explained by the second harmonic generation process.
In order to achieve the deepest modulation with minimum RAM, the 
working frequency has been chosen at $\Omega_{\rm m}/2 \pi= 660$ kHz.

\section{Results}

The green laser output  was successfully locked to the FP cavity. A 
cavity transmission of 7 \% was obtained  and the measured finesse 
was 45\,000. The frequency
locking  was stable for durations up to several hours. Figure 4 shows 
the spectral density 
of the difference between the laser frequency and the instantaneous 
resonance 
frequency of the cavity. 
This spectral density has an overall minimum value of $\sim$1 
mHz/$\sqrt{\rm Hz}$, while 
it stays below 10 mHz/$\sqrt{\rm Hz}$ in the region 1-200 Hz 
(excluding peaks due to
the 50 Hz mains frequency and its harmonics). This result is 
comparable again to 
the one obtained using the IR output of the NPRO laser. The estimated 
value of the frequency shift due to RAM is $\sim$ 0.1 Hz. Finally, table 2 
below summarizes a few relevant characteristics and measured 
performances of the system.

\section{Conclusions}

It has been shown that it is possible to lock the green output of a 
doubled IR NPRO laser to a resonant, high-finesse, Fabry Perot 
cavity, using an originally modified form of the Pound-Drever-Hall 
technique. In particular, 
phase modulation of the beam was achieved using the piezoelectric 
transducer acting on the infrared pump laser crystal. This allows for 
a 
simple locking scheme involving a previously unavailable 
wavelength (532 nm in this case), which is now in the visible domain.

\section*{Acknowledgements}

This work has been carried out within the PVLAS collaboration: the 
authors 
wish to thank all the members of the group.

\newpage

\section*{References}
\begin{enumerate}
\item
E. Zavattini, F. Brandi, M. Bregant, G. Cantatore, S. Carusotto,
F. Della Valle, G. Di Domenico, U. Gastaldi, E. Milotti, R. Pengo,
G. Petrucci, E. Polacco, G. Ruoso, G. Zavattini,
{\it The PVLAS Collaboration: Experimental Search For 
Anisotropy Of The Phase Velocity Of Light In Vacuum Due To 
A Static Magnetic Field}, in Proc. of QED2000 - G. Cantatore Ed. - 
AIP Conf. Proceedings {bf 564}, 77 (2001);

D. Bakalov, F. Brandi, G. Cantatore, G. Carugno, S. Carusotto,
F. Della Valle, A. M. De Riva, U. Gastaldi, E. Iacopini, P. Micossi,
E. Milotti, R. Onofrio, R. Pengo, F. Perrone, G. Petrucci, E. Polacco,
C. Rizzo, G. Ruoso, E. Zavattini and G. Zavattini, Quant. and Semiclass. Opt. {\bf 10}, 239 (1998);

PVLAS homepage: http://www.ts.infn.it/experiments/pvlas.

\item
A. Abramovici, W. E. Althouse, R.W.P. Drever, Y.  Gursel, S. Kawamura, F. J.  Raab, D. Shoemaker,  L. Sievers, R.E.  Spero, K.S. Thorne, R.E. Vogt, R. Weiss,  S.E. Whitcomb, M.E. Zucker, Science {\bf 256}, 325 (1992);

C. Bradaschia {\it et al.}, Nucl. Instrum. Meth. A {\bf 289}, 518 
(1990).
\item
L. Conti, M. De Rosa, F. Marin,
Appl. Optics {\bf 39}, 5732 (2000);

L. Conti, M. Cerdonio, L. Taffarello, J.P.  Zendri, A. Ortolan, C.  Rizzo, G. Ruoso, G.A. Prodi, S. Vitale, G.  Cantatore, E. Zavattini,
Rev. Sci. Instrum. {\bf 69}, 554 (1998). 

\item
U. Gastaldi and M. Placentino, Nucl. Instrum. Meth. A {\bf 451}, 318 (2000).
\item
R. V. Pound, Rev. Sci. Instrum. {\bf 17}, 490 (1946);

R. W. P. Drever, J.L. Hall, F.V. Kowalsky, J. Hough, G. M. Ford, A.J. Munley, and H. Ward, Appl. Phys. B {\bf 31}, 97 (1983).

\item
G. Cantatore, F. Della Valle, E. Milotti, P. Pace, F. Perrone, E. Polacco, C. Rizzo, G. Ruoso, E. Zavattini, G. Zavattini,
Rev.   Sci. Instrum. {\bf 66}, 2785 (1995).

\item
Model Prometheus, Innolight Gmbh, Hannover, Germany (http://www.innolight.de).

\item
Made by Research Electro Optics, Boulder, CO, U.S.A. (http://www.reoinc.com). 

\end{enumerate}

\newpage

FIGURE CAPTIONS

\noindent
Figure 1: Experimental set-up.

\noindent
Figure 2:
Residual amplitude modulation (RAM) as a function of the modulation 
frequency. The signal on the PZT has an amplitude of 48 mV$_{\rm 
pp}$.

\noindent
Figure 3:
Index of modulation versus the voltage $V_{\rm p}$ applied to the 
piezoelectric transducer at the chosen modulation frequency of 660 
kHz.

\noindent
Figure 4: Spectral density 
of the difference between the laser frequency and the resonance 
frequency of the cavity. This is obtained by measuring the noise 
spectrum at the error point of figure 1 and by multiplying it by the 
slope of the error signal  in the same point. The large peaks are due 
to 50 Hz and its harmonics.

\newpage

\begin{table}[tbp]
	\centering
	\caption{\footnotesize RAM and $\beta$ as a function of the voltage 
applied to the 
	PZT for some values of the frequency.}
	\begin{tabular}{|l|l|l|l|}
	\hline
	$\nu$ (kHz) & $V_{\rm pp}/\beta$ (mV) & RAM/$V_{\rm pp}$ 
	(mV$^{-1}$) & RAM/$\beta$ \\
	\hline
	600 &	109 &	1.97$\times 10^{-6}$ &	2.15$\times 10^{-4}$ \\
	660	&	100 &	1.12$\times 10^{-6}$ &	1.12$\times 10^{-4}$  \\
    733	&	104 &	2.21$\times 10^{-6}$ &	2.31$\times 10^{-4}$ \\
    830 &	163 &	4.06$\times 10^{-6}$ &	6.60$\times 10^{-4}$ \\
    900	&	93  &	1.06$\times 10^{-5}$ &	9.84$\times 10^{-4}$ \\
    \hline
	\end{tabular}
	\label{}
\end{table}

\begin{table}[tbp]
 \centering	\caption{\footnotesize Summary of 
relevant system characteristics and measured performances}	
\begin{tabular}{|l|l|}	 \hline  
laser wavelength & 532 nm \\
laser output power @532 nm & 100 mW \\
   	FP cavity length & 87 cm \\
   	FP cavity finesse & 45000 \\
    FP cavity Q factor  & 1.4 $\times 10^{11}$ \\
    beam waist at cavity center & 0.7 mm \\
    light power density at cavity center & 120 W/cm$^{2}$	\\
    energy stored in the cavity & 50 nJ \\
    NPRO temperature control BW & 1 Hz \\ 
NPRO temperature control dynamic range & 60 GHz \\
NPRO PZT control BW & 100 kHz\\
NPRO PZT control dynamic range & 200 MHz \\
\hline
\end{tabular}
\label{cara}
\end{table}

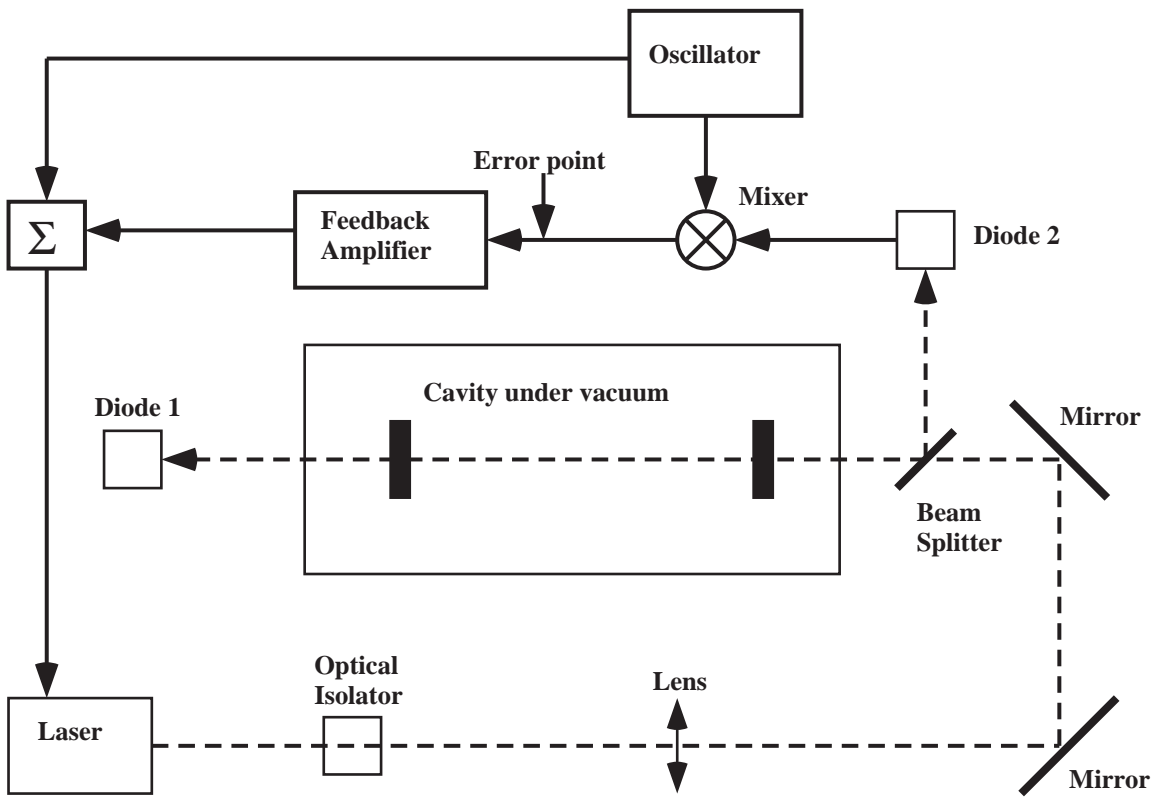
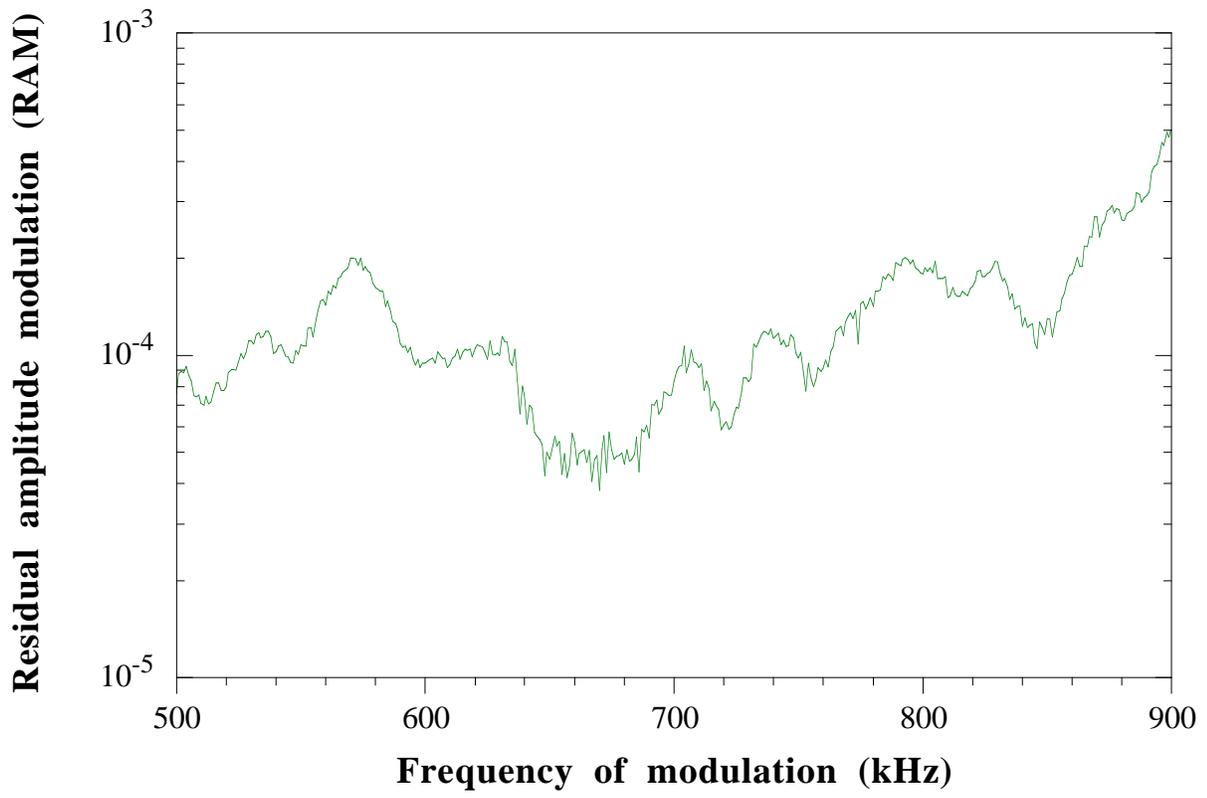
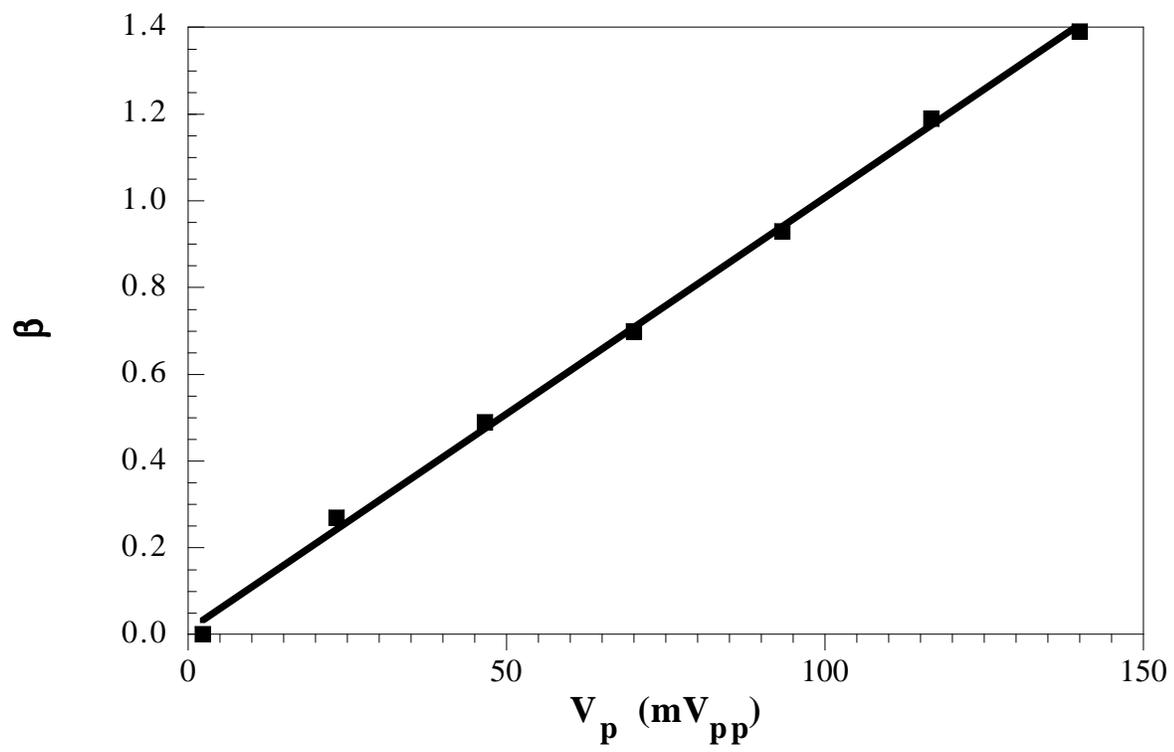
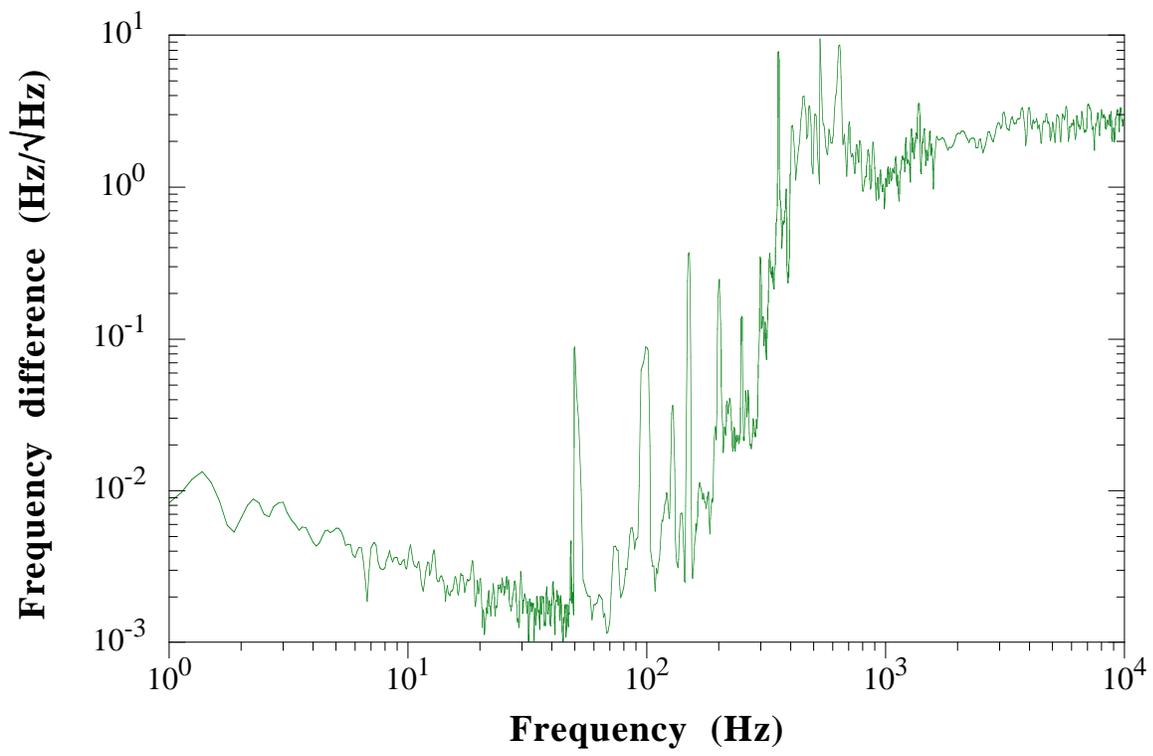
\end{document}